\documentclass[aps,prd,secnumarabic,nobibnotes,twocolumn,superscriptaddress]{revtex4-1}
\usepackage{amsfonts}
\usepackage{mathrsfs}
\usepackage{amsmath}
\usepackage{color}
\usepackage{natbib}
\usepackage{graphicx}
\usepackage{bm}
\usepackage{amssymb}
\usepackage{xspace}
\usepackage{epstopdf}
\usepackage{dcolumn}
\usepackage{multirow}
\usepackage[colorlinks=true, letterpaper=true, pdfstartview=FitV, linkcolor=blue, citecolor=blue, urlcolor=blue]{hyperref}
\usepackage{wrapfig}
\usepackage{booktabs}

\makeatletter

\newcommand{\Rmnum}[1]{\expandafter\@slowromancap\romannumeral #1@}
\makeatother

\begin{document}
\title{Accurate recipe for predicting valley Linear Weyl phonons in two dimensions}

\author{Mingmin Zhong}\thanks{M.Z., H.L., and J.W. contributed equally to this manuscript.}
\address{School of Physical Science and Technology, Southwest University, Chongqing 400715, China}

\author{Haibo Liu}\thanks{M.Z., H.L., and J.W. contributed equally to this manuscript.}
\address{School of Physical Science and Technology, Southwest University, Chongqing 400715, China}

\author{Jianhua Wang}\thanks{M.Z., H.L., and J.W. contributed equally to this manuscript.}
\address{School of Physical Science and Technology, Southwest University, Chongqing 400715, China}
\author{Chengwu Xie}
\address{School of Physical Science and Technology, Southwest University, Chongqing 400715, China}
\author{Hongkuan Yuan}
\address{School of Physical Science and Technology, Southwest University, Chongqing 400715, China}

\author{Zeying Zhang}\thanks{Corresponding authors}\email{zzy@mail.buct.edu.cn}
\address{College of Mathematics and Physics, Beijing University of Chemical Technology, Beijing 100029, China}

\author{Xiaotian Wang}\thanks{Corresponding authors}\email{xiaotianwang@swu.edu.cn}
\address{School of Physical Science and Technology, Southwest University, Chongqing 400715, China}
\address{Institute for Superconducting and Electronic Materials (ISEM), University of Wollongong, Wollongong 2500, Australia}

\author{Gang Zhang}\thanks{Corresponding authors}\email{zhangg@ihpc.a-star.edu.sg}
\address{Institute of High Performance Computing, Agency for Science, Technology and Research (A*STAR), 138632, Singapore}

\begin{abstract}
The discovery of topological quantum states in two-dimensional (2D) systems is one of the most promising advancements in condensed matter physics. Linear Weyl point (LWP) phonons have been theoretically investigated in some 2D materials. Especially, Jin, Wang, and Xu [Nano Lett. 2018, 18, 12, 7755-7760] proposed in 2018 that the candidates with threefold rotational symmetry at the corners of the hexagonal Brillouin zone can host LWP phonons with a quantized valley Berry phase. However, all the candidates with hexagonal lattices may not host LWP phonons at $K$ ($K'$) high-symmetry points (HSPs). Hence, a more accurate recipe for LWP phonons in 2D is highly required. This work provides an exhaustive list of valley LWP phonons at HSPs in 2D by searching the entire 80 layer groups (LGs). We found that the valley LWP phonons can be obtained at HSPs in 11 of the 80 LGs. Guided by the symmetry analysis, we also contributed to realizing the ideal 2D material with valley LWP phonons. We identified the existence of the valley LWP phonons in eleven 2D material candidates with 11 LGs. This work offers a method to search for valley LWPs in 2D phononic systems and proposes 2D material candidates to obtain the valley LWP phonons.

\end{abstract}
\maketitle

\section{Introduction}

Phonons are energy quanta of lattice vibrations. They contribute significantly to several physical properties, including thermal conductivity, superconductivity, thermoelectricity, and specific heat. Like the topological electronic nature, the field of topological phononics can be established by introducing the crucial theorems and concepts of topology into the study of phonons~\cite{add1,add2,add3,add4,add5}. Topological phonons in solid materials are connected to specific atomic lattice vibrations that typically fall within a terahertz frequency range, providing a rich platform for studying various boson-related quasiparticles~\cite{add6,add7,add8,add9,add10,add11,add12,add13,add14,add15,add16,add17,add18,add19}. Hence, the search for materials containing topological phonons has become a priority in topological physics.

The study of topological phonons in condensed matter systems has gained popularity over the past five years. Furthermore, novel topological phonons, including nodal point~\cite{add20,add21,add22,add23,add24,add25,add26,add27,add28,add29}, nodal line~\cite{add30,add31,add32,add33,add34,add35,add36,add37,add38,add39}, and nodal surface phonons~\cite{add39,add40,add41,add42,add43}, are generally discovered in three-dimensional (3D) real materials. Among them, a series of nodal point phonons---with different topological charges, numbers of degeneracies, order of dispersion around the point, and types of slopes of crossing bands forming the point---were predicted in theory using first-principle calculations and symmetry analysis. For example, the linear Weyl point (LWP) has relativistic linear dispersion in any direction in momentum space. It is topologically protected; it can exist in 3D crystals without any space group symmetry (except translation symmetry). Normally, LWP can be categorized as type-I or type-II depending on the dispersion of two crossing bands~\cite{add44,add45,add46,add47,add48}. Wang \textit{et al}.~\cite{add18} in 2019 proposed the appearance of symmetry-protected ideal Type-II LWP phonons in zinc-blende CdTe in theory. The LWPs are located along the high-symmetry lines at the boundaries of the fcc Brillouin zone (BZ) due to the absence of the spin-orbital coupling (SOC) effect. The phonon surface arcs connecting WPs with opposite chirality are guaranteed to be extremely long; moreover, they are readily observable in experiments. In the same year, ideal type-II LWP phonons were also reported in wurtzite CuI by Liu \textit{et al}.~\cite{add49} through first-principles investigations. In 2021, You, Sheng, and Su~\cite{add50} proposed the simultaneous presence of type I and type II LWP phonons in $T$-carbon---a newly discovered carbon allotrope.

Note that the twofold degenerate LWPs have also been proposed in a small number of two-dimensional (2D) phononic systems~\cite{add5,add51}. Compared to 3D materials, 2D materials with less symmetrical constraints may more intuitively display the clean characteristics of topological phonons~\cite{add52,add53}. In 2018, Jin, Wang, and Xu~\cite{add5} presented a recipe for LWP phonon states with quantized valley Berry phase in 2D hexagonal lattices through first-principles calculations. They~\cite{add5} reported that the candidates possessing the three-fold rotational symmetry at the corners of the hexagonal Brillouin zone host valley LWP phonons. Furthermore, they~\cite{add5} also selected monolayers CrI$_3$ and YGaI as examples to demonstrate the topologically nontrivial characteristics of LWP phonons. Subsequently, Li \textit{et al}.~\cite{add51} in 2020 proposed that 2D graphene hosts four types of LWP phonons and a nodal ring phonon in its phonon dispersion.

\begin{figure}[tbp]
\includegraphics[width=8.5cm]{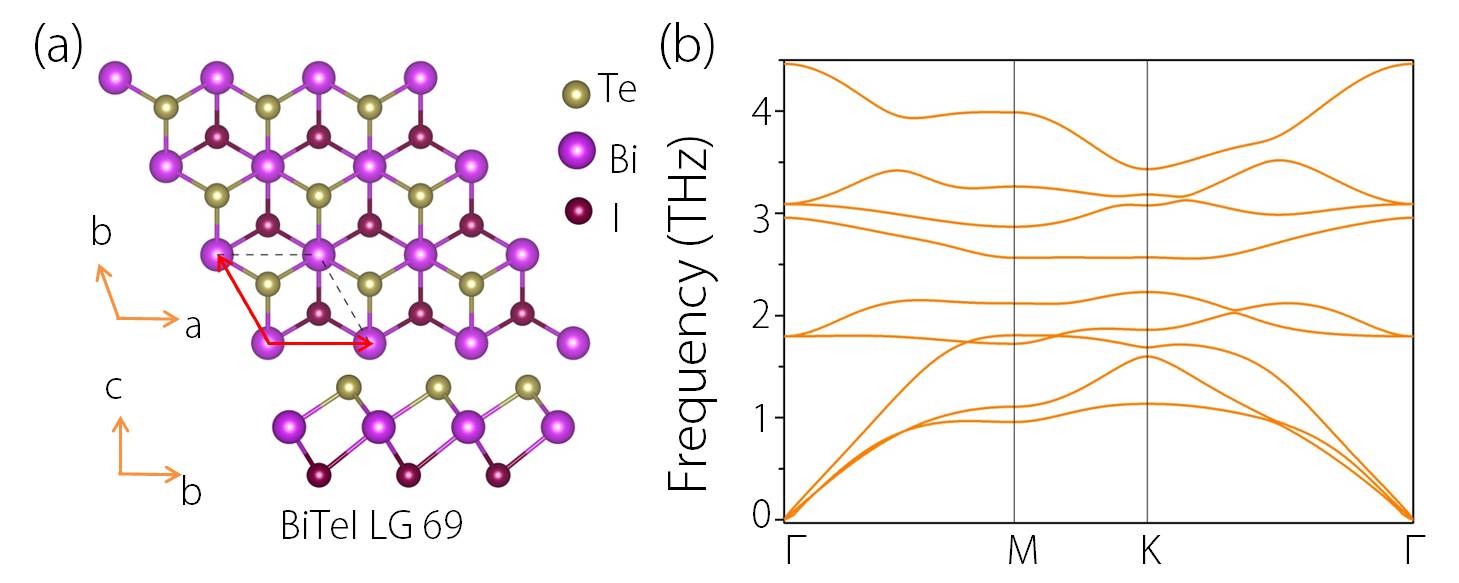}
\caption{(a) Structural model of 2D BiTeI with LG 69. (b) Phonon dispersion of BiTeI monolayer. A 3 $\times$ 3 $\times$ 1 supercell of the monolayer was adopted to calculate force constants.
\label{fig1}}
\end{figure}

Consider the little co-group $G_k^0$ of $K$ point in a hexagonal lattice. Two possible cases exist for $G_k^0$ to be an abelian group, i.e., $C_3$ and $C_{3h}$. Nonlinear irreducible representation exists for the non-abelian group corresponding to the enforced Weyl point at $K$ point. The irreducible representations of abelian groups are always linear, meaning that the band structure cannot degenerate. However, Herring's rule was used to learn that two conjugate representations of $C_3$ are mostly stuck tighter by the combination of lattice symmetry and $\mathcal{T}$ considering time-reversal symmetry $\mathcal{T}$. There are some exceptions, such as $K$ point of LG 69, the co-representation type of $K_2$ in LG 69 can be calculated as follow:

\begin{equation}
\begin{split}
\frac{1}{\vert P \vert} \sum _\alpha \chi(\alpha ^2) & =\frac{1}{3} (\chi(\sigma _{v1}^2)+\chi(\sigma _{v2}^2)+\chi(\sigma _{v3}^2)) \\
& =\frac{1}{3} (1+1+1)=1,
\end{split}
\end{equation} where $\chi$ is the character of $K_2$, $\alpha$ is the symmetry operator that satisfies $\alpha K = -K\ mod\ g$, and $g$ is a reciprocal lattice vector. That is the co-representation type of $K_2$ is type (a) [see Eq. 1 of Ref.~\cite{add54} and there is no additional degeneracy. Then, we selected a 2D BiTeI (with LG 69) monolayer as an example to achieve clarity. The structural model for the 2D monolayer with hexagonal lattices is shown in Fig.~\ref{fig1}(a). The phonon dispersion for 2D BiTeI monolayer is shown in Fig.~\ref{fig1}(b). The figure shows that the material did not host valley LWPs at $K$ and $K'$ high-symmetry points (HSPs). Specifically, the phonon branches did not cross at $K$ and $K'$ HSPs even though having the three-fold rotational ($C_3$) symmetry.

Hence, a more accurate recipe for predicting valley LWP phonons in 2D is still required besides the recipe proposed by Jin, Wang, and Xu~\cite{add5}. We searched all 80 LGs with time-reversal symmetry $\mathcal{T}$ and theoretically identified 11 LG candidates that host LWP phonons at HSPs $K$ and $K'$(see Table ~\ref{table1}) to systematically identify all possible LWP phonons at HSPs for each layer group (LG). Moreover, we also contributed to obtaining the material for the LWP phonons at HSPs. We identified the presence of LWP phonons in eleven 2D candidate materials with eleven LGs---2D AlSiTe$_3$ with LG 66, 2D GaSe$_2$O$_8$ with LG 68, 2D ScPS$_3$ with LG 70, 2D InSiTe$_3$ with LG 71, 2D ZrSe$_2$ with LG 72, 2D C$_{10}$F$_3$H$_3$ with LG 73, 2D p-MoS$_2$ with LG 75, 2D Bismuthylene with LG 76, 2D As$_2$O$_3$ with LG 77, 2D C$_5$N with LG 79, and 2D AuBe$_2$ with LG 80---using first-principle calculations.

\section{Symmetry analysis}

Table ~\ref{table1} shows that LGs 66, 68, 70-73,75-80 can host LWP phonons at HSPs. We selected one LG (LG 77) to understand the existence of LWP at $K$ ($K'$) HSP. The existence of LWP other LGs (66, 68, 70, 71, 72, 73, 75, 76, 79, and 80) was explained by constructing the $k \cdot p$ model around $K$ (see the Supplemental Material (SM)~\cite{add57}).

\begin{table}[hb]
  \centering
  \caption{LG candidates (and their corresponding space groups [SGs]) that can host LWPs at HSPs in 2D phononic systems. This table also includes the locations of the LWP phonons, the correspondence generators and labels associated with the LWPs, and the 2D material candidates.}
  \label{table1}
  \begin{tabular}{@{\extracolsep{\fill}}ccccccc}
    \hline
    \hline
    LG  &       SG   &   SG/LG  & Generator   & Label & Materials  \\
	No. &       No.  &  symbol &     \     &   \      &  \    &  \       \\
	\hline
    66  &     147 & $\textit{P}\overline{3}$    & $C_3^+$,$\mathcal{IT}$&    $K_2$$K_3$  &  AlSiTe$_3$  \\
    68  &     150 & $\textit{P}321      $       & $C_3^+$,$C_{21}^{''}$ &    $K_3$       &  GaSe$_2$O$_8$  \\
    70  &     157 & $\textit{P}31m      $       & $C_3^+$,$\sigma_{d1}$ &    $K_3$       &  ScPS$_3$  \\
    71  &     162 & $\textit{P}\overline{3}1m$  & $C_3^+$,$\sigma_{d1}$,$\mathcal{IT}$ &    $K_3$       &  InSiTe$_3$  \\
    72  &     164 & $\textit{P}\overline{3}m1$  & $C_3^+$,$C_{21}^{''}$,$\mathcal{IT}$ &    $K_3$       & ZrSe$_2$  \\
    73  &     168 & \textit{P}6                 & $C_3^+$,$C_{6}^{+} \mathcal{T}$  &    $K_2$$K_3$  &  C$_{10}$F$_3$H$_3$  \\
    75  &     175 & $\textit{P}6/m     $        & $C_3^+$,$S_{3}^{-}$,$\mathcal{IT}$ &    $K_2$$K_3$/$K_5$$K_6$ &  p-MoS$_2$  \\
    76  &     177 & $\textit{P}622     $        & $C_3^+$,$C_{21}^{''}$,$C_{6}^{+} \mathcal{T}$   &    $K_3$       &  Bismuthylene  \\
    77  &     183 & $\textit{P}6mm     $        & $C_3^+$,$\sigma_{d1}$,$C_{6}^{+} \mathcal{T}$   &    $K_3$       & As$_2$O$_3$ \\
    79  &     189 & $\textit{P}\overline{6}2m$  & $C_3^+$,$C_{21}^{''}$,$S_{3}^{-}$ &    $K_5$/$K_6$ &  C$_5$N  \\
    80  &     191 & $\textit{P}6/\textit{mmm}$  & $C_3^+$,$C_{21}^{''}$,$S_{3}^{-}$,$\mathcal{IT}$  &    $K_5$/$K_6$ &  AuBe$_2$  \\

     \hline
    \hline
    \end{tabular}
\end{table}

For LG 77, the existence of LWP can be understood as follows: Consider the $K_3$ co-irreducible representation of $K$ for LG 77. The representation matrix for $K_3$ can then be written as~\cite{add55}:
\begin{equation}\label{2}
\begin{split}
{\cal{C}} _{3} ^+ =\begin{pmatrix} -\frac{1}{2} & -\frac{\sqrt{3}}{2} \\
 \frac{\sqrt{3}}{2} & -\frac{1}{2} \end{pmatrix}, \sigma _{d1} =\begin{pmatrix} 1 & 0 \\   0 & -1 \end{pmatrix},
{\cal{C}} _{6} ^+ \mathcal{T} =\begin{pmatrix} \frac{1}{2} & -\frac{\sqrt{3}}{2} \\  \frac{\sqrt{3}}{2} & \frac{1}{2} \end{pmatrix}.
\end{split}
\end{equation}

Based on the representation matrix, using the MagneticKP package~\cite{add56}, we can then construct the $k \cdot p$ model around $K$:
\begin{equation}\label{3}
{\cal{H}} _{\rm 77}=\quad\begin{pmatrix}{\cal{C}}_{0,1}+k_y {\cal{C}}_{1,1} & k_x {\cal{C}}_{1,1} \\  k_x {\cal{C}}_{1,1} & {\cal{C}}_{0,1}-k_y {\cal{C}}_{1,1} \end{pmatrix},
\end{equation} which is indeed an LWP at $K$ HSP.

\begin{figure*}
\includegraphics[width=13cm]{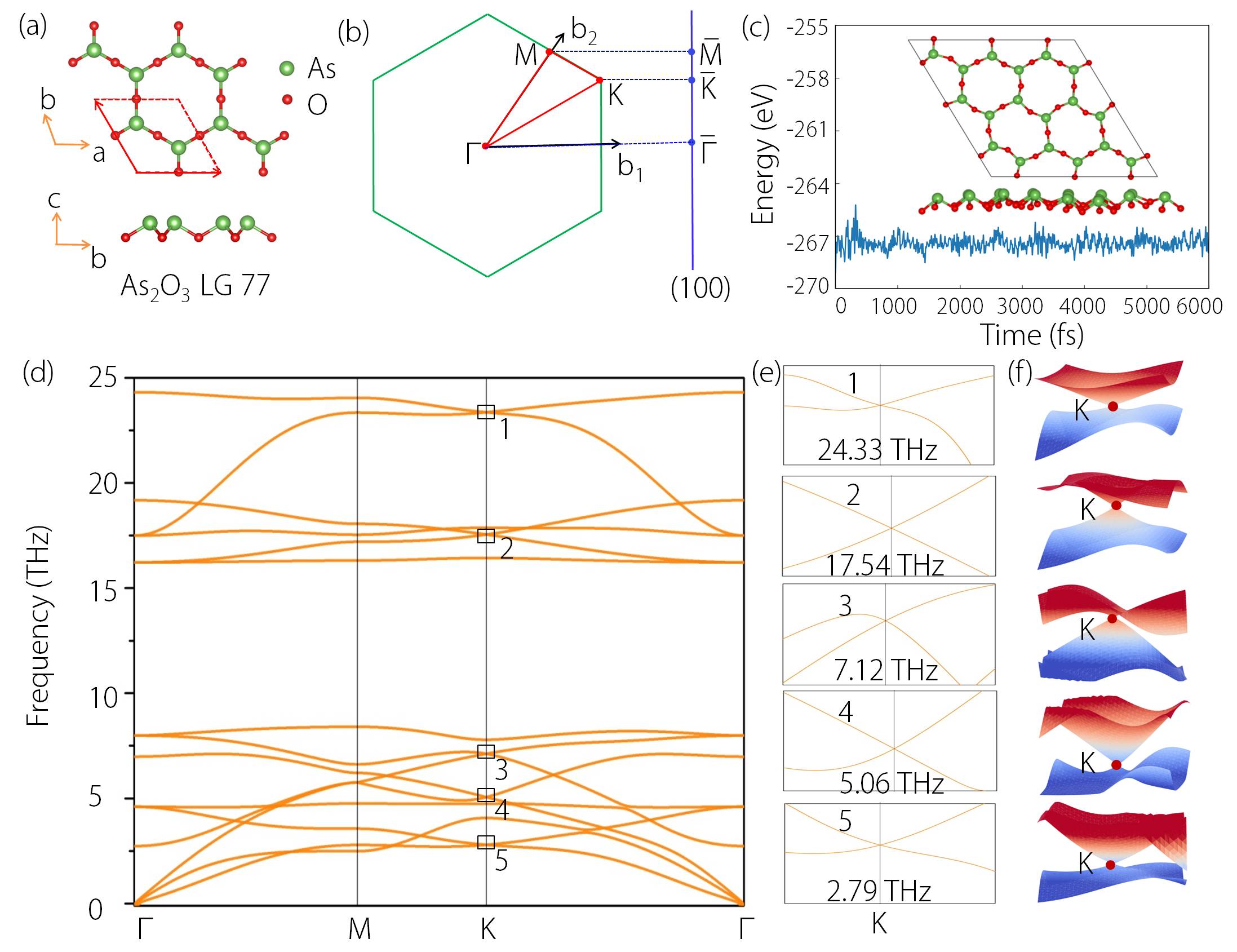}
\caption{(a) Structural model of As$_2$O$_3$ with LG 77. (b) The 2D BZ and its projection to the (100) edge. (c) Total potential energy fluctuation of (3 $\times$ 3 $\times$ 1) supercell for As$_2$O$_3$ monolayer during AIMD simulations at 300 K. The inset in (c) indicates the snapshot at the end of simulation of 6000 fs. (d) Phonon dispersion of 2D As$_2$O$_3$ monolayer along the high-symmetry paths $\Gamma$--M--K--$\Gamma$. The LWPs (labeled with Nos. 1-5) located at different frequencies are marked by black boxes. (e) Enlarged phonon bands around the 5 LWP phonons. (f) 3D plots of the phonon bands around the LWPs 1-5 at HSP $K$.
\label{fig2}}
\end{figure*}

\section{2D material examples with valley LWP phonons}

Eleven 2D material candidates with LGs 66, 68, 70-73,75-80, including 2D AlSiTe$_3$, 2D GaSe$_2$O$_8$, 2D ScPS$_3$, 2D InSiTe$_3$, 2D ZrSe$_2$, 2D C$_{10}$F$_3$H$_3$, 2D p-MoS$_2$, 2D Bismuthylene, 2D As$_2$O$_3$, 2D C$_5$N, and 2D AuBe$_2$, are presented in this work to support our symmetry analysis.

We selected 2D As$_2$O$_3$ with LG 77 as a typical example to discuss the occurrence of the LWP phonons at HSPs. The results of the other ten 2D materials can be referred to the SM~\cite{add57}.

The structural model of 2D As$_2$O$_3$ can be found in Materials Cloud two-dimensional crystals database (MC2D)~\cite{add64}. 2D As$_2$O$_3$ monolayer hosts LG 77; its relaxed lattice constants are a = b = 5.378 \AA. Fig.~\ref{fig2}(a) shows the structural model, in which As and O atoms occupied 2$b$ (0.333, 0.666, 0.529) and 3$c$ (0.000, 0.500, 0.482) Wyckoff positions, respectively. We performed an ab initio molecular dynamic (AIMD) simulation~\cite{add65} of a large supercell (3 $\times$ 3 $\times$ 1) with a Nos{\'e}-Hoover thermostat~\cite{add66} at 300 K to test the thermal stability of the As$_2$O$_3$ monolayer. Fig.~\ref{fig2}(c) shows the results of the fluctuations in the potential energy as a function of simulation time at 300 K. No structural destruction of the As$_2$O$_3$ monolayer after 6000 fs at a time step of 1 fs (see the structure in the inset) except for thermally-induced fluctuations, indicating that the As$_2$O$_3$ monolayer at room temperature is thermally stable. We calculated the phonon dispersion of the As$_2$O$_3$ monolayer along the high-symmetry paths $\Gamma$--M--K--$\Gamma$ (see Fig.~\ref{fig2}(b)) using density functional perturbation theory~\cite{add67}; the results are shown in Fig.~\ref{fig2}(d). A 3 $\times$ 3 $\times$ 1 supercell was adopted to calculate force constants. Furthermore, 15 branches in the phonon dispersion consisting of 3 acoustic and 12 optic branches were found owing to having five atoms in the primitive cell of the 2D As$_2$O$_3$ monolayer. Fig.~\ref{fig2}(d) displays a lack of imaginary frequency in the phonon spectrum, implying the dynamic stability of monolayer As$_2$O$_3$. More importantly, a series of crossing points (labeled as 1-5) exist at HSP $K$ (or $K'$) that agree well with the symmetry analysis in Table ~\ref{table1}. The crossing point 1 (approximately 24.33 THz) was extremely clean, i.e., the two bands forming the crossing point did not overlap with the other bands around 24 THz. For clarity, the enlarged crossing points 1-5 at HSP $K$ and the 3D plots of crossing points 1-5 are shown in Fig.~\ref{fig2}(e) and Fig.~\ref{fig2}(f), respectively. Note that clean LWP phonons were easily detectable and crucial for topological quantum phonon transport applications. The crossing points 1-5 were twofold degenerate Weyl points with linear phonon dispersion.

\begin{figure}
\includegraphics[width=8.5cm]{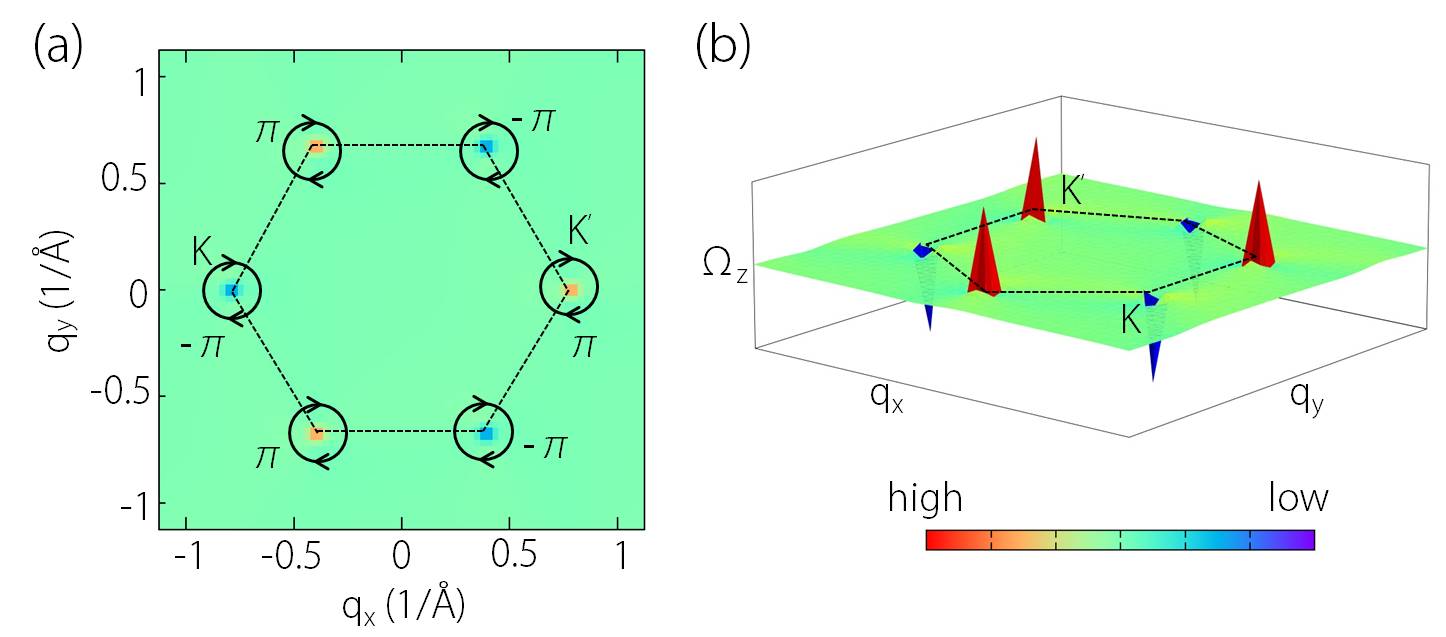}
\caption{Top (a) and side (b) views of Berry curvature distributions for LWP 1. We used dotted lines to show the first BZ. In (a), the phonon Berry phases around $K$ and $K'$ valleys are nontrivial and quantized with values of $-\pi$ and $\pi$, respectively.
\label{fig3}}
\end{figure}

Selecting LWP 1 as an example, we plotted the distributions of Berry curvature $\mathbf{\Omega}_z(\boldsymbol{q})$ in momentum space. The phonon Berry curvature can reveal the topological features of LWP phonons at HSP $K$ (or $K'$). The results from the top and side views of Berry curvature distributions are shown in Fig.~\ref{fig3}(a) and Fig.~\ref{fig3}(b). Nonzero $\mathbf{\Omega}_z(\boldsymbol{q})$ diverged at $K$ (or $K'$) valleys and disappeared away from these valleys. Furthermore, the $K$ and $K'$ valleys hosted opposite $\mathbf{\Omega}_z(\boldsymbol{q})$ (see Fig.~\ref{fig3}(b)). Therefore, the integral of $\mathbf{\Omega}_z(\boldsymbol{q})$ in the entire BZ must be zero. As shown in Fig.~\ref{fig3}(a), we calculated the Berry phases of the loop encircling the LWP at $K$ and $K'$. The results showed that the Berry phases around $K$ and $K'$ valleys were nontrivial and quantized.

\begin{figure}[b]
\includegraphics[width=8.5cm]{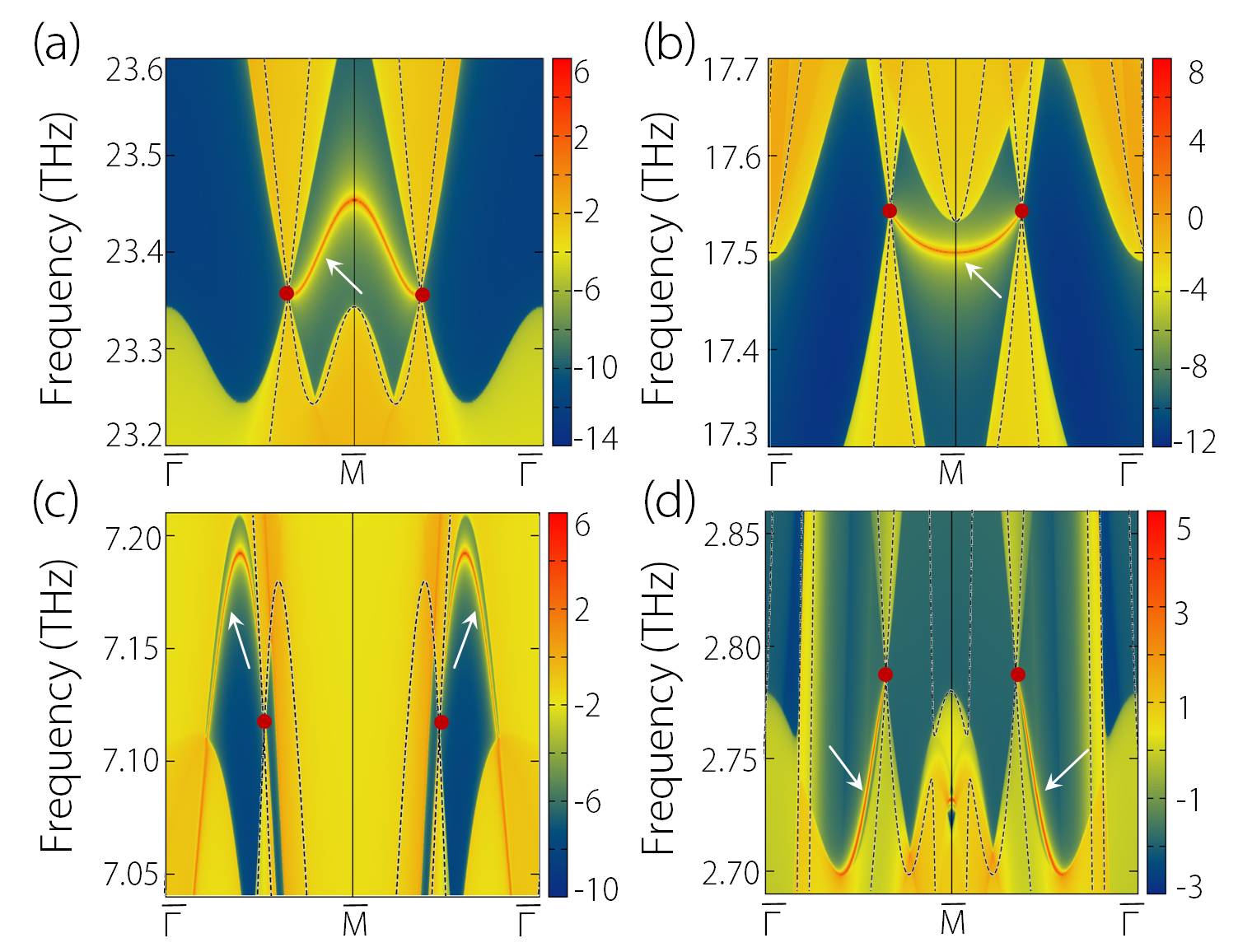}
\caption{(a)-(d) Edge states arising from the projections of the LWPs 1, 2, 3, and 5, respectively. The dotted lines in (a)-(d) show the phonon dispersions around the LWPs 1, 2, 3, and 5 (marked by red dots), respectively. The white arrows in (a)-(d) show the edge states originating from the projections of the LWPs with Nos. 1, 2, 3, and 5.
\label{fig4}}
\end{figure}

LWP phonons with nontrivial and quantized Berry phases lead to a nontrivial topological edge state~\cite{add68,add69,add70,add71,add72,add73}. We calculated the edge states for the LWPs in the 2D As$_2$O$_3$ monolayer along (100) direction using the imaginary parts of the Green's function within the WannierTools code~\cite{add74}. The local density of states (LDOS) of phonons around LWPs 1, 2, 3, and 5 are shown in Fig.~\ref{fig4}. Phononic edge states, arising from the projections of LWPs, are visible, as excepted.

Besides the 2D As$_2$O$_3$ monolayer with LG 77, the remaining ten 2D materials with LGs 66, 68, 70--73,75, 76, 78--80 were included in SM~\cite{add57}, where the visible LWP phonons at HSPs ($K$ and $K'$) and phononic edge states could also be observed to support the illuminating results of the topological phonons in 2D.

We show the structural models of 2D AlSiTe$_3$ with LG 66, 2D GaSe$_2$O$_8$ with LG 68, 2D ScPS$_3$ with LG 70, 2D InSiTe$_3$ with LG 71, 2D ZrSe$_2$ with LG 72, 2D C$_{10}$F$_3$H$_3$ with LG 73, 2D p-MoS$_2$ with LG 75, 2D Bismuthylene with LG 76, 2D C$_5$N with LG 79, and 2D AuBe$_2$ with LG 80 in Fig. S1, Fig. S4, Fig. S7, Fig. S10, Fig. S13, Fig. S16, Fig. S19, Fig. S21, Fig. S24, Fig. S27, and Fig. S30, respectively (see SM~\cite{add57}). Among them, 2D AlSiTe$_3$, 2D GaSe$_2$O$_8$, 2D InSiTe$_3$, 2D ZrSe$_2$ were searched from the MC2D~\cite{add75,add76,add77,add78}. Furthermore, 2D ScPS$_3$ was screened from the computational 2D materials database~\cite{add79}. 2D C$_{10}$F$_3$H$_3$ can be created by substituting N in C$_7$N$_3$~\cite{add80} with C, on which one F atom and one H atom are adsorbed. 2D C$_5$N~\cite{add81} was obtained using a mechanical exfoliation approach from 3D carbon-rich nitride C$_5$N. Moreover, P-MoS$_2$ was proposed by Lin \textit{et al}.~\cite{add82} in 2016. It has a significantly different crystal structure than H-MoS$_2$~\cite{add83}. Crucially, it has a 27$\%$ larger specific surface area, making it a superior contender for the energy industry and catalysts. Bismuthylene~\cite{add84}, the porous allotrope of a bismuth monolayer, was predicted by Zhang \textit{et al}. in 2017 to be a 2D topological insulator using first-principles calculations. Furthermore, AuBe$_2$ monolayer with planar hexacoordinate s-block metal atoms was predicted by Wang \textit{et al}.~\cite{add85} in 2022 to be a superconducting global minimum Dirac material with two perfect Dirac node-loops.

The phonon dispersions for AlSiTe$_3$, GaSe$_2$O$_8$, ScPS$_3$, InSiTe$_3$, ZrSe$_2$,  C$_{10}$F$_3$H$_3$, p-MoS$_2$, Bismuthylene, C$_5$N, and AuBe$_2$ along $\Gamma$--M--K--$\Gamma$ paths are shown in Fig. S2, Fig. S5, Fig. S8, Fig. S11, Fig. S14, Fig. S17, Fig. S20, Fig. S23, Fig. S26, and Fig. S29, respectively (see SM~\cite{add57}), in which the multiple valley LWPs at HSPs $K$ (or $K'$) are visible. Like 2D As$_2$O$_3$ in Fig.~\ref{fig4}, the Berry curvatures have an extremum at $K$ and $K'$ for all cases but with opposing signs. However, they are strictly zero at all other points, reflecting that the valley LWPs at $K$ and $K'$ have opposite charges.

As examples, the edge states related to a series LWPs are shown in Fig. S3, Fig. S6, Fig. S9, Fig. S12, Fig. S15, Fig. S18, Fig. S21, Fig. S24, Fig. S27, and Fig. S30, respectively (see SM~\cite{add57}). The lack of overlap between topological edge states and bulk states allows surface-sensitive instruments to easily detect the phononic edge states. Given that several interesting physical properties are associated with phonons, the above-mentioned 2D materials would be an ideal candidate for elucidating the fundamental physical phenomena associated with topological phonons in 2D.

\section{Summary}

In conclusion, we studied the symmetry conditions of 80 LGs and discovered that valley WLP phonons could appear at HSPs $K$ and $K'$ in 11 of the 80 LGs. After that, we predicted eleven 2D materials hosting twofold degenerate LWP phonons at HSPs. We also determined the LWPs at $K$ and $K'$ valleys to have an opposite quantized Berry phase, implying that they are topologically nontrivial. Moreover, the phonon edge states, arising from the projections of LWPs at $K$ (or $K'$), were visible. This work provides a more accurate recipe for predicting the twofold degenerate LWPs in 2D phononic systems. Furthermore, this work can help investigate other types of emergent particles in 2D phononic systems.

\section*{Acknowledgments}

This work was supported by the the National Natural Science Foundation of China (No. 12004028), and the Natural Science Foundation of Chongqing (No. CSTB2022NSCQ-MSX0283).

\end{document}